\documentclass[prd,twocolumn,superscriptaddress,floatfix,amsmath,amssymb]{revtex4-1}
\usepackage[english]{babel}
\usepackage{graphicx}
\usepackage{dcolumn}
\usepackage{bm}
\usepackage{verbatim}
\usepackage{mathrsfs}
\usepackage{cancel}
\usepackage{color}
\newcommand{\ket}[1]{\left| {#1} \right\rangle}

\newcommand{\ro}{r_{\Omega}}
\newcommand{\eq}[1]{(\ref{#1})}

\newcommand{\qr}{{q_\text{R}}}
\newcommand{\ql}{{q_\text{L}}}

\def\slashchar#1{\setbox0=\hbox{$#1$} 
\dimen0=\wd0 
\setbox1=\hbox{/} \dimen1=\wd1 
\ifdim\dimen0>\dimen1 
\rlap{\hbox to \dimen0{\hfil/\hfil}} 
#1 
\else 
\rlap{\hbox to \dimen1{\hfil$#1$\hfil}} 
/ 
\fi}

\begin{document}

\title{Non-monotonic entanglement of physical EM field states in non-inertial frames}
\author{Miguel Montero}
\affiliation{Instituto de F\'{i}sica Fundamental, CSIC, Serrano 113-B, 28006 Madrid, Spain}
\author{Marco del Rey}
\affiliation{Instituto de F\'{i}sica Fundamental, CSIC, Serrano 113-B, 28006 Madrid, Spain}
\author{Eduardo Mart\'{i}n-Mart\'{i}nez}
\affiliation{Instituto de F\'{i}sica Fundamental, CSIC, Serrano 113-B, 28006 Madrid, Spain}
\affiliation{Institute for Quantum Computing, Department of Physics and Astronomy and Department of Applied Mathematics, University of Waterloo, 200 University
Avenue W, Waterloo, Ontario, N2L 3G1, Canada}

\begin{abstract}
We develop a general technique to analyse the quantum effects of acceleration on realistic spatially-localised electromagnetic field states entangled in the polarization degree of freedom. We show that for this setting, quantum entanglement may build up as the acceleration increases, providing a clear signature of the quantum effects of relativistic acceleration.\end{abstract}

\maketitle

\section{Introduction}
One of the most widely known results in relativistic quantum information is the notion that acceleration may have non-trivial effects on entanglement. A number of works have studied this issue through transformations between inertial and accelerated Fock bases (among many others, \cite{Alicefalls,AlsingSchul,Edu4,Edu9}). Previous results considered tailored families of states which greatly simplified calculations but whose physical interpretation was not clear. Recently, \cite{Dravette} introduced a projective detector model which constitutes a promising new approach to the issue of field entanglement in non-inertial frames. It was shown in \cite{Dravette} that such detectors present the expected thermal response to the inertial vacuum state of the field (i.e. the Unruh effect \cite{Unruh0}) and the model provides a good effective description of particle detection.  
In this paper, we will use this projective model as a practical method to study field entanglement for localised two single-photon bipartite electromagnetic field states.

There are, however, two different approaches to use the projective model introduced in \cite{Dravette}. Here, we will develop a technique to analyse states by working out the Bogoliubov transformations and the change of basis inertial-accelerated modes in some approximate but physically feasible scenario. A different technique, that allows to obtain exact results for Gaussian states without going through the Bogoliubov coefficients calculation, is being developed and will appear elsewhere \cite{Inprepar}.
 
While the latter can be very handy to analyse  squeezed states such as those produced in parametric down conversion, it is not clear to what extent it can be used to analyse two single-photon states as those studied in this paper. Due to the relevance of non-monotonic entanglement behaviour reported here in the two single-photon regime, it is worth exploring the first approach. Also, the computation of the Bogoliubov transformations between localised inertial and accelerated Fock bases is of much interest in itself for its possible future use.

Let us use the detector model  mentioned above \cite{Dravette} to explore bipartite entanglement for two single-photon states entangled in helicities with a Gaussian spread in frequencies. This  spatially localised system is observed by two partners: an inertial one, Alice, looking at one of the photons, and an accelerated one, Rob, observing the other photon.

The practical limiting factor of the general formalism presented in this paper is computational complexity, which may grow very quickly for some cases of interest. We study for which physical regimes results can be given within current computing power, and suggest how the computational issues of the formalism may be overcome beyond that.

\section{Setting}
We will consider two different observers, Alice and Rob. They are interested in studying field correlations in the electromagnetic field \footnote{As usual in the literature, to analyse these effects more clearly we will assume that the acceleration lies in the polarisation quantisation axis so no Thomas precession occurs.}. Alice is an inertial observer, while Rob undergoes a motion with constant proper acceleration $\tilde{a}$ and acceleration frequency $a=\tilde{a}/c$.

Rob's trajectory is best described in terms of Rindler coordinates. We introduce two sets of Rindler coordinates $(\xi_\pm,\tau_\pm)$ whose relation with the Minkowskian coordinates $(x,t)$ is
\begin{align}\label{change}
ct=&\pm\xi_\pm \sinh\left(a\tau_\pm\right),\qquad x=\pm\xi_\pm\cosh\left(a\tau_\pm\right),
\end{align}
with $x>|t|$ for $(\xi_+,\tau_+)$   and $x<-|t|$ for $(\xi_-,\tau_-)$. These coordinates naturally define two globally hyperbolic and causally disconnected submanifolds in flat spacetime, each being the mirror image of the other, which we call regions I and II following the standard notation of \cite{Alicefalls,Edu9}. Any field theory in flat spacetime can be regarded as two independent field theories, one in each of these regions \cite{Birrell}. Without loss of generality we place Rob in  Region I. No operations will be carried out involving measurement or communication with region II. Although we make use of some mathematical constructions  which mix region I and region II operators, we do so merely as a convenient computational tool. Our results are independent of whatever happens in the causally disconnected region II, thus, the entanglement we find is of a different nature to that present in the Minkowski vacuum state, which displays correlations between regions I and II \cite{Unruh0}.

At this point, there are two paths to go through. On one hand, we might take a specific  model for Alice's and Rob's detectors (such as Unruh-DeWitt),  prepare the detectors in a particular state and then let the field-detectors system evolve. After some time the detectors would lose and gain entanglement due to the non-trivial effects of Rob's acceleration. Recent examples of this approach are found in \cite{Andrzej, Ralphy, Beilok}. 

Still, this method has the drawback of being dramatically dependent on the details of the detector model assumed and  it is very difficult to explore beyond first order perturbation theory or inside small cavities. Besides, depending on the model, the detectors may not inherit all the field correlations and may develop further correlations owing to the specific form of the interaction which do not come directly from field entanglement.

 On the other hand, field entanglement has been successfully studied for decades by means of projective measurements on the field state, independently of any detector model. Entanglement between correlated photon pairs is usually accessed this way \cite{Milburn_computes}. This approach also avoids all the complications associated with the detector-dependent settings previously discussed. This is the original approach used  (among others) in \cite{Alicefalls,AlsingSchul,Edu4,Edu9,Mig1} with completely delocalised Unruh modes \cite{Edu9}. In this line,  using the localised projective detector formalism developed in \cite{Dravette} we do not face any problem coming from non-locality and acceleration dependence associated to the Unruh modes.

\section{Procedure}

As common in the literature \cite{Alicefalls,Mig3}, the Fock space for Alice will be constructed in terms of Minkowski creation operators $a^\dagger_{\omega,\sigma}$ acting on the Minkowski vacuum state $\ket{0}_\mathcal{M}$, defined by
\begin{align}a_{\omega,\sigma}\ket{0}_\mathcal{M}=0.\end{align} 
Here, $\sigma\in\{\uparrow,\downarrow\}$ denotes helicity.

Rob, as an accelerated observer, will build his Fock basis by means of Rindler modes. In this case, the Fock space will be constructed in terms of Rindler creation operators $a^\dagger_{\Omega,\sigma,\text{I}}$ (and their complement $a^\dagger_{\Omega,\sigma,\text{II}}$) acting on the Rindler vacuum state $\ket{0}_\mathcal{R}$, defined by
\begin{align}a_{\Omega,\sigma,\text{I}}\ket{0}_\mathcal{R}=a_{\Omega,\sigma,\text{II}}\ket{0}_\mathcal{R}=0.\end{align}
 The Rindler modes are labeled by their dimensionless Rindler frequency $\Omega\equiv\Omega'/a$, where $\Omega'$ is the Rindler frequency.

Let us consider that Alice and Rob carry detectors capable of exploring a particular set of modes of the field, but without assuming anything about the particular detector model. Rob's associated field vacuum is not $\ket{0}_\mathcal{M}$, but rather, the Rindler vacuum $\ket{0}_\mathcal{R}$. This means that Rob will be able to make projective measurements in the subspace spanned by the basis 
 \begin{align}\mathcal{B}=\left\{ \frac{1}{\sqrt{n!}}(d_{\uparrow}^\dagger)^n \ket{0}_\mathcal{R},\frac{1}{\sqrt{n!}}(d_{\downarrow}^\dagger)^n \ket{0}_\mathcal{R}\vert n\in \mathbb{N}\right\}\label{cawen}\end{align}
 with the operators $d_{\sigma}^\dagger$, which create one `detector mode' excitation, being a linear combination of Rindler creation operators of definite helicity,
 \begin{align}\label{thetas}d_{\sigma}^\dagger=\int_0^\infty d\Omega\,g(\Omega) a^\dagger_{\Omega,\sigma , \text{I}}.\end{align}
 
We will analyse a very general family of arbitrarily spatially localised entangled states of the form
 \begin{align}\label{state}\ket{\Psi}=P\ket{a}_\text{A}\ket{x}_\text{Rob}+Q\ket{b}_\text{A}\ket{y}_\text{Rob},\qquad \vert P\vert^2+\vert Q\vert^2=1.\end{align}
 Here, the states $\ket{x}_\text{Rob}$ and $\ket{y}_\text{Rob}$ will be a pair of Minkowskian wavepacket one-particle excitations of opposite helicities, i.e.
 \begin{align}\ket{x}_\text{Rob}&= \left(\int d\omega\ x(\omega) a^\dagger_{\omega,\uparrow}\right)\ket{0}_\mathcal{M},\nonumber\\ \ket{y}_\text{Rob}&= \left(\int d\omega\ y(\omega) a^\dagger_{\omega,\downarrow} \right)\ket{0}_\mathcal{M}.\end{align}
$\ket{a}_\text{A}$ and $\ket{b}_\text{A}$ can be assumed to be a pair of Minkowskian wavepackets similar to $\ket{x}_\text{Rob}$ and $\ket{y}_\text{Rob}$.
 
As Rob will probe his part of the field state by means of projective measurements on the basis \eqref{cawen}, to compute the effect of these measurements we will have to express first $\ket{x}_\text{Rob}$ and $\ket{y}_\text{Rob}$ in the Rindler basis. 
 
This change of basis is most easily computed via an intermediate change to the so-called Unruh modes. For bosonic fields, the Unruh modes are defined in terms of the Rindler modes by
 \begin{align}\label{umodes}a_{\text{R},\Omega,\sigma}&=\cosh\ro a_{\text{I},\Omega,\sigma}-\sinh\ro a^\dagger_{\text{II},\Omega,-\sigma},\nonumber\\
a_{\text{L},\Omega,\sigma}&=\cosh\ro a_{\text{II},\Omega,-\sigma}-\sinh\ro a^\dagger_{\text{I},\Omega,\sigma},\end{align}
with $\tanh\ro=e^{-\pi\Omega}$.

The advantage of these modes is that  the Minkowski vacuum factorises as \cite{Birrell} 
\begin{align}\ket{0}_\mathcal{M}=\bigotimes_\Omega \ket{0}_{\Omega,\uparrow}\ket{0}_{\Omega,\downarrow}\label{vac}\end{align}
where $a_{\text{R},\Omega,\sigma}\ket{0}_{\Omega,\sigma}=a_{\text{L},\Omega,\sigma}\ket{0}_{\Omega,\sigma}=0$. The explicit form of $\ket{0}_{\Omega,\sigma}$ in terms of Rindler modes has been found elsewhere \cite{Mig3}. Eq. \eq{vac} implies that the Unruh and Minkowski modes share the same vacuum state, this meaning that Minkowski-Unruh change of basis preserves the number of particles. Therefore, we may express $\ket{x}_\text{Rob}$ and $\ket{y}_\text{Rob}$ as
\begin{align}
\ket{x}_\text{Rob}&=\int d\Omega\alpha^\text{R}_{\omega\Omega} x(\omega)\ket{\uparrow_{\text{R}\Omega}}+\int d\Omega\alpha^\text{L}_{\omega\Omega} x(\omega) \ket{\uparrow_{\text{L}\Omega}},\nonumber\\
\ket{y}_\text{Rob}&=\int d\Omega\alpha^\text{R}_{\omega\Omega} y(\omega)\ket{\downarrow_{\text{R}\Omega}}+\int d\Omega\alpha^\text{L}_{\omega\Omega} y(\omega) \ket{\downarrow_{\text{L}\Omega}},\nonumber\\
\text{with} &\ket{\sigma_{\text{X}\Omega}}=\label{xyrin} a^\dagger_{\text{X},\Omega,\sigma}\ket{0}_\mathcal{M},\ \text{X}=\{\text{L},\text{R}\} 
\end{align}
where the $\alpha^\text{X}_{\omega\Omega}$ are the coefficients of the relevant change of basis, which are computed for a scalar field in \cite{Edu9} and adapted for an electromagnetic field in \cite{Mig3}.

Substituting \eq{xyrin} in \eq{state}, we finally express $\ket{\Psi}$ as a linear combination of states entangled between Alice's modes and Unruh modes of the form $\ket{\sigma 	\qr}_\Omega=(q_\text{L}a^\dagger_{\text{L},\Omega,\sigma}+q_\text{R}a^\dagger_{\text{R},\Omega,\sigma})\ket{0}_\mathcal{M}$,  ($\ql$ being such that $|\qr|^2+|\ql|^2=1, \ \ \qr\geq \ql$). Namely,
\begin{align}\label{bast}\ket{\Psi}&=\int d\Omega\  \ket{\Phi}_\Omega,\quad \ket{\Phi}_\Omega=\ket{\phi}_\Omega \bigotimes_{\Omega'\neq\Omega}\ket{0}_\Omega,\\\ket{\phi}_\Omega&=P_\Omega\ket{a}_\text{A}\ket{\uparrow\qr_1}_{\stackrel{\!\!\!\!\!\!\!\!\Omega}{\text{Rob}}}+Q_\Omega\ket{b}_\text{A}\ket{\downarrow\qr_2}_{\stackrel{\!\!\!\!\!\!\!\!\Omega}{\text{Rob}}}.\label{fff}\end{align}
These states $\ket{\phi}_\Omega$ are precisely those studied in previous works on field entanglement in non-inertial frames \cite{Edu9,Edu10,Mig1}, and their form in the Rindler basis is well known. We remark that, in contrast to previous works \cite{Alicefalls,AlsingSchul,Edu9,Mig1}, the state \eqref{state} (and consequently \eqref{bast}) is the same no matter the acceleration of Rob.

Now, we recall that Rob's detector probes the non-monochromatic modes \eqref{thetas}. To build a complete basis of the Fock space, we need to complete \eqref{thetas} with their orthogonal complement $D_\perp=\{\alpha^\dagger_1,\alpha^\dagger_2\ldots\}$ where $\alpha^\dagger_i$ are one-particle creation operators so that $\mathcal{S}=\{d_{\uparrow}^\dagger,d_{\downarrow}^\dagger,\alpha^\dagger_1,\alpha^\dagger_2\ldots\}$ is complete. In this basis, for a state $\ket{n_1\ n_2\ldots}$ the detector will only be sensitive to the first two entries, which correspond to the detector modes. All we need to do then is tracing out the irrelevant set of modes $\alpha_1,\ldots\alpha_n$. We will end up with a reduced field state containing all the field entanglement accessible to our detector. At this point, we may quantify this entanglement through any suitable entanglement measure, such as the negativity \cite{Negat}.

Note that the previous results reported in \cite{Alicefalls,AlsingSchul,Edu9,Mig1} correspond to the choice of $\ket{x}_\text{Rob}$ and $\ket{y}_\text{Rob}$ to be Unruh excitations of a single fixed frequency $\Omega_\text{field}$, and imposing a detector profile in adimensional frequencies $g(\Omega)=\delta(\Omega-\Omega_\text{det})$. In order to probe the entanglement for those simple modes, these  works used a  single-frequency detector with fixed $\Omega_\text{det}'$, which would couple to a dimensionless Rindler frequency $\Omega_\text{det}=\Omega_\text{det}'/a$. That means that the $\Omega$ really probed by the detector depends on its acceleration. For each acceleration $a$, the field state was chosen peaked around a certain dimensionless Rindler frequency  $\Omega_\text{field}(a) $, so as to always have $\Omega_\text{field} =\Omega'_\text{det}/a$. In contrast, using the projective detector model \cite{Dravette} we can probe the same  acceleration-independent field state.

{\it Peaked detectors.--}
With the formalism above, we may consider a physical  state and look at the behaviour of entanglement with acceleration on a peaked distribution of Rindler frequencies. Although we will discuss below that our formalism can be extended to arbitrarily spatially smeared detectors, this is a most reasonable first step towards the study of realistic experimental scenarios. Approximately single-frequency detector modes have been constructed in \cite{Edu9} and constitute a nice and simple starting point. Such a detector mode will be spread in space with some finite characteristic length dependent on the frequency spread. 

Let us consider the state \eq{state} with $P=Q=1/\sqrt{2}$ and Gaussian mode profiles, 
\begin{align}x(\omega)=y(\omega)=(2\pi\omega)^{-1/4}\ e^{-\frac{(\omega-\omega_0)^2}{4\sigma^2}}.\label{gpf}\end{align}
If we choose $\ket{a}_\text{A}$ and $\ket{b}_\text{A}$ as another pair of Gaussian modes centered at a frequency far from $\omega_0$, then \eq{state} represents a normalised version of a two single-photon field state maximally entangled in polarisations.  

We now introduce a realistic approximation for a single-mode detector for Rob by caracterising its spectral decomposition in terms of Rindler modes as
\begin{equation}
g_\text{det} (\Omega')=  \Delta{\Omega'}_\text{det}^{-1/2} \Pi\left(\frac{\Omega'-\Omega'_\text{det}}{\Delta\Omega'_\text{det}}\right)
\end{equation}
where $\Pi(x)$ is the unit step function, the characteristic function of the $[-\frac12,\frac12]$ interval. A plot of this frequency profile is shown in Fig. 1.
\begin{figure}[hbtp]
\begin{center} 
\includegraphics[width=.46\textwidth]{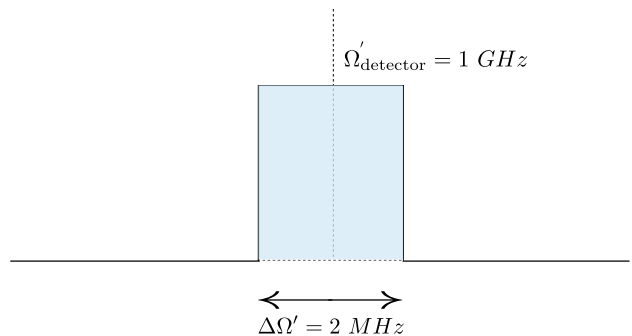}
\caption{Frequency profile of the simplified detector mode considered in the text.}
\label{detect}
\end{center}
\end{figure}

After  tracing out the unobserved field frequencies through eqs. \eq{bast} and \eq{fff}, we compute the negativity of the relevant reduced state as a function of the acceleration. In order to make this calculation simpler we assume that  
\begin{align} r_{\Omega_\text{det}-\Delta\Omega_\text{det}/2}\approx r_{\Omega_\text{det}}  \approx r_{\Omega_\text{det}+\Delta\Omega_\text{det}/2}. \label{condition}
\end{align}
From $r_\Omega$ definition \eq{umodes} we find that for \eq{condition} to hold one must have
\begin{align}
\frac{\Delta r_{\Omega_\text{det}}}{r_{\Omega_\text{det}}}  \approx &\Big|\frac{d \ln r_\Omega}{d\Omega} \Big|_{\Omega_\text{det}} \nonumber \Delta\Omega_\text{det}= \\	&\frac{\pi e^{-\pi\Omega_\text{det}}\Delta \Omega_\text{det} }{\text{atanh}(e^{-\pi\Omega_\text{det}})(1+e^{-2\pi\Omega_\text{det}})}\ll 0.1
\end{align}
which assuming a quality factor $Q=\Omega_\text{det}/\Delta\Omega_\text{det}=500$ (the typical value for microwave filters) happens to be valid  for $\Omega_\text{det}=\Omega_\text{det}'/a \lesssim 10 $. Considering $\Omega'_\text{det}=1$ GHz, that implies we must have $\tilde a = ac \gtrsim 3 \cdot 10^{16} m/s^{2}$. In this regime, we may assure that the observed effects are the direct consequence of Rob's acceleration, without any other effects playing any important role.

\begin{figure}[hbtp] 
\begin{center} 
\includegraphics[width=.46\textwidth]{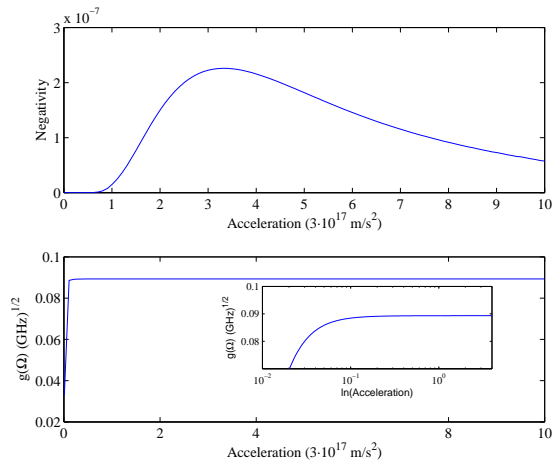}
\caption{(a) Negativity as a function of Rob's acceleration for the field state \eq{state} with Gaussian profiles \eq{gpf} with $\sigma=0.01$  GHz, $\omega_0= 1$ GHz. The Rindler spread of the detector mode is $\lambda=2$ MHz. (b) Rindler spread of the field state, defined in \eq{xyrin}, as a function of Rob's acceleration via the map $\Omega=\Omega'_d/a$, with detector Rindler dimensionful frequency $\Omega'_d=1$ GHz. The inset provides a logarithmic scale of the profile.}
\label{graf}
\end{center} 
\end{figure}

Figure \ref{graf} shows our results for a reasonable choice of the relevant parameters, along with the Rindler spread of the state (the coefficients of the left and right excitations in \eq{xyrin}). 

Remarkably, the entanglement amplification phenomenon first reported in \cite{Mig1} still survives in this more physical scenario, and in fact it is present in single-photon entangled states.

Note that the amount of entanglement shown in Fig \ref{graf} is very small. This responds only to the fact that, in this example, we are considering a ultra-narrow-band detector: When looked at from the accelerated frame, the localised Minkowskian states spread over a broad Rindler frequency spectrum. In the simple example presented here we analyse correlations only for a highly peaked frequency spectrum of Rindler frequencies, so a great amount of the entanglement is lost due to the detector not seeing all the relevant frequencies.

\begin{figure}[hbtp] 
\begin{center} 
\includegraphics[width=.46\textwidth]{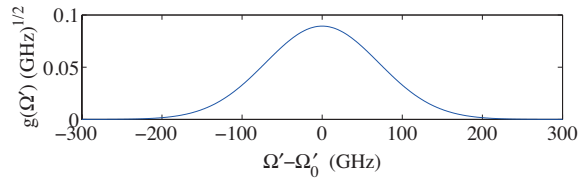}
\caption{Localised Rindler frequency spread of the Gaussian-localised Minkowskian state \eq{gpf} for $\omega_0=1$ Ghz and $\sigma=0.01$ Ghz, for  $\tilde{a}=3\cdot10^{17}\ m/s^2$. A detector mode with this profile in Rindler frequencies would be maximally tuned to the field state being probed and therefore would result in much higher entanglements. $\Omega_0$ comes from an appropriate phase choice of the Minkowskian state.}
\label{migueloceronte}
\end{center} 
\end{figure}

In order to see more of the entanglement of the field state, we need a detector whose bandwidth is the most similar to the Rindler frequency distribution corresponding to the initial Minkowski wavepacket. Figures \ref{graf}b and \ref{migueloceronte} show  how a Minkowksi Gaussian wavepacket transforms into a well localised state in terms of Rindler modes. If the bandwidth of the detector approaches this localised distribution of Rindler modes, the amount of entanglement detected will be much higher. As we will discuss below, considering wider-band detectors in this formalism is straightforward, but it has a price in terms of computational complexity.

In any case our results mean that protocols of entanglement distillation \cite{Distil}  could be implemented to detect entanglement generation due to acceleration and therefore provide an unmistakable witness of the Unruh effect, easier to detect than the entanglement degradation reported in \cite{Alicefalls,Edu9} and others.

\section{Broadband detectors}
Our formalism can be applied to more general  detector  profiles and multimode detection: Once we have determined both the state and the detector mode(s), we may use \eq{bast} to find the relevant reduced state of detector modes. We discuss in this section how computational difficulties appear when considering a more general detector with an arbitrary large mode spread and how to overcome them. 

Note that \eq{bast} for the field state prior to tracing out the unobserved degrees of freedom includes a continuous product of frequencies. This means that to numerically obtain a reduced state from it, we need to make a sampling in frequencies, and then take a limit of small discretisation step. Let us discretise the frequencies so as to have $m$ sample points within the spread of $g(\Omega)$ and allow up to $n$ excitations per frequency. Then a lower bound in the computational complexity of obtaining the reduced state can be derived just by computing the number of individual operations required to manage the sparse matrices involved in the calculation, giving a complexity growing as
\begin{align}\label{LB}\mathcal{O}[a(n)^m].\end{align}
where $a(n)$ is a monotonously increasing function of $n$ satisfying $a(n)\geq 5$. Such a huge lower bound gives an idea of the intractability of the problem.

Until a way around the computational problems of a large number of frequencies is found, the only procedure to study these regimes is through quantum simulations and condensed matter analog systems (see e.g. \cite{MarcoEdu}) or by considering field states with convenient properties, such as Gaussian states as done in \cite{Dravette} and other families which will  be treated elsewhere \cite{Inprepar}. 

\section{Conclusions}

We have devised a scheme to analyse field entanglement in non-inertial frames for arbitrary single-photon field states and detector frequency response. To do this we have used the projective detector model introduced in \cite{Dravette}. As a particular case, we dealt with reasonable electromagnetic field states: We have analysed entanglement behaviour of  a two-mode photon state entangled in helicities. We have shown that the quantum effects of relativistic acceleration can actually amplify entanglement and not only destroy it.

Entanglement amplification phenomena have been reported before for some rather unphysical families of states \cite{Mig1}, but the present work clearly shows that the effect is a genuine consequence of acceleration in less idealised field states.

This formalism allows us to consider very general and realistic states.  In particular, we have thoroughly analysed the case of peaked detectors and studied the rapidly scaling computational costs of considering wide-band detection. These difficulties may be overcome through the use of quantum simulations: Instead of using analog systems to test predictions, we may use them to make predictions.  Also, all the conclusions are exportable to a a static black hole scenario by means of the formalism developed in \cite{Edu7}.

{\it Acknowledgments.--} We thank Juan Le\' on for interesting discussions and Carlos Sab\'{i}n for his very helpful comments. We also thank Andrzej Dragan and I. Fuentes for the enlightening discussions about previous works on the topic. M. dR. was supported by a CSIC JAE-PREDOC grant and by QUITEMAD S2009-ESP-1594.

%
\end{document}